\documentclass[twocolumn]{aastex631}

\usepackage{hyperref}

\accepted{version}

\submitjournal{ApJL; accepted 2022 November 27; published 2022 December 19}

\shorttitle{SN 2021ocs: extreme stripping of progenitor O-Mg layer}
\shortauthors{Kuncarayakti et al.}
\graphicspath{{./}{figures/}}

\begin{document}


\title{Late-time H/He-poor circumstellar interaction in the type-Ic supernova SN~2021ocs: an exposed oxygen-magnesium layer and extreme stripping of the progenitor\footnote{Based on observations collected at the European Organisation for Astronomical Research in the Southern Hemisphere under ESO programme 108.2282.001.}}

\author[0000-0002-1132-1366]{H. Kuncarayakti}
\affiliation{Tuorla Observatory, Department of Physics and Astronomy, FI-20014 University of Turku, Finland}
\affiliation{Finnish Centre for Astronomy with ESO (FINCA), FI-20014 University of Turku, Finland}

\author[0000-0003-2611-7269]{K. Maeda}
\affiliation{Department of Astronomy, Graduate School of Science, Kyoto University, Sakyo-ku, Kyoto 606-8502, Japan}

\author{L. Dessart}
\affiliation{Institut d’Astrophysique de Paris, CNRS-Sorbonne Universit\'e, 98 bis boulevard Arago, F-75014 Paris, France}

\author{T. Nagao}
\affiliation{Tuorla Observatory, Department of Physics and Astronomy, FI-20014 University of Turku, Finland}

\author{M. Fulton}
\affiliation{Astrophysics Research Centre, School of Mathematics and Physics, Queen's University Belfast, Belfast, BT7 1NN, UK}

\author{C. P. Guti\'errez}
\affiliation{Finnish Centre for Astronomy with ESO (FINCA), FI-20014 University of Turku, Finland}
\affiliation{Tuorla Observatory, Department of Physics and Astronomy, FI-20014 University of Turku, Finland}

\author[0000-0003-1059-9603]{M. E. Huber}
\affiliation{Institute for Astronomy, University of Hawaii, 2680 Woodlawn Drive, Honolulu HI 96822, USA}

\author{D. R. Young}
\affiliation{Astrophysics Research Centre, School of Mathematics and Physics, Queen's University Belfast, Belfast, BT7 1NN, UK}

\author{R. Kotak}
\affiliation{Tuorla Observatory, Department of Physics and Astronomy, FI-20014 University of Turku, Finland}

\author{S. Mattila}
\affiliation{Tuorla Observatory, Department of Physics and Astronomy, FI-20014 University of Turku, Finland}
\affiliation{School of Sciences, European University Cyprus, Diogenes street, Engomi, 1516 Nicosia, Cyprus}

\author[0000-0003-0227-3451]{J.~P. Anderson}
\affiliation{European Southern Observatory, Alonso de C\'ordova 3107, Casilla 19, Santiago, Chile}
\affiliation{Millennium Institute of Astrophysics MAS, Nuncio Monsenor Sotero Sanz 100, Off. 104, Providencia, Santiago, Chile}

\author{L. Ferrari}
\affiliation{Facultad de Ciencias Astron\'omicas y Geof\'isicas, Universidad Nacional de La Plata, Paseo del Bosque S/N, B1900FWA La Plata, Argentina}
\affiliation{Instituto de Astrof\'isica de La Plata (IALP), CONICET, Argentina}

\author{G. Folatelli}
\affiliation{Facultad de Ciencias Astron\'omicas y Geof\'isicas, Universidad Nacional de La Plata, Paseo del Bosque S/N, B1900FWA La Plata, Argentina}
\affiliation{Instituto de Astrof\'isica de La Plata (IALP), CONICET, Argentina}
\affiliation{Kavli Institute for the Physics and Mathematics of the Universe (WPI), The University of Tokyo, 5-1-5 Kashiwanoha, Kashiwa, Chiba 277-8583, Japan}

\author[0000-0003-1015-5367]{H. Gao}
\affiliation{Institute for Astronomy, University of Hawaii, 2680 Woodlawn Drive, Honolulu HI 96822, USA}

\author[0000-0002-7965-2815]{E. Magnier}
\affiliation{Institute for Astronomy, University of Hawaii, 2680 Woodlawn Drive, Honolulu HI 96822, USA}

\author{K. W. Smith}
\affiliation{Astrophysics Research Centre, School of Mathematics and Physics, Queen's University Belfast, Belfast, BT7 1NN, UK}

\author[0000-0003-4524-6883]{S. Srivastav}
\affiliation{Astrophysics Research Centre, School of Mathematics and Physics, Queen's University Belfast, Belfast, BT7 1NN, UK}

%
%
%
%



\begin{abstract}

Supernova (SN) 2021ocs was discovered in the galaxy {NGC~7828 ($z = 0.01911$)} within the interacting system Arp~144, and subsequently classified as a normal type-Ic SN around peak brightness. VLT/FORS2 observations in the nebular phase at {148 d} reveal that the spectrum is dominated by oxygen and magnesium emission lines of different transitions and ionization states: O I, [O I], [O II], [O III], Mg~I, and Mg~II. Such a spectrum has no counterpart in the literature, though it bears a few features similar to those of some interacting type Ibn and Icn SNe. Additionally, SN 2021ocs showed a blue color, $(g-r) \lesssim -0.5$~mag, after the peak {and up to late phases}, atypical for a type-Ic SN. Together with the nebular spectrum, this suggests that SN 2021ocs underwent late-time interaction with an H/He-poor circumstellar medium (CSM), resulting from the pre-SN progenitor mass loss during its final $\sim$1000 days. 
The strong O and Mg lines and the absence of strong C and He lines suggest that the progenitor star's O-Mg layer is exposed, which places SN~2021ocs as the most extreme case of massive progenitor star's envelope stripping in interacting SNe, followed by type-Icn (stripped C-O layer) and Ibn (stripped He-rich layer) SNe. This is the first time such {a} case is reported in the literature.
{SN 2021ocs emphasizes the importance of late-time spectroscopy of SNe, even for those classified as normal events, to reveal the inner ejecta and progenitor star's CSM and mass loss}.


\end{abstract}



\section{Introduction} \label{sec:intro}

Stripped-envelope supernovae (SESNe) {include} a broad variety of {events}, all of which show little or no hydrogen in the spectrum. SNe type Ic are deficient of H and He, hence the progenitors are traditionally thought to be Wolf-Rayet (WR) stars \citep[e.g.][]{woosley95}.
The progenitor stars of SESNe require significant mass loss in order to remove the outer H envelope. Stellar winds, binary interaction, and eruptions are among the prominent mechanisms \citep[see e.g.][for a review]{smith14}.
A highly massive single star ($M_{\rm ZAMS} \gtrsim25~M_\odot$) is required to form a WR star through wind stripping. This casts doubt if there is sufficient number of WR stars to explain the observed rate of SESNe, which implies that a significant number of SESNe must have come from lower-mass stars in close binary systems \citep[e.g.][]{yoon10,smith11}.

The SN progenitor mass loss may result in a circumstellar medium (CSM), which may interact with the SN ejecta after the explosion. SNe interacting with H-free CSM are rare, nevertheless modern surveys have discovered a sample of them, which are now classified into types Ibn (He-rich CSM) and Icn (H/He-poor CSM). Some of these objects are thought to be explosions of genuine massive WR stars {\citep{pastorello07,smith17,galyam22}, or less massive stars \citep{sanders13,dessart22,pellegrino22,davis22}, within a dense CSM}.

SN 2021ocs was discovered by the Asteroid Terrestrial-impact Last Alert System (ATLAS) survey \citep{tonry18,smith20} on 2021-05-30 (UTC time used throughout) as ATLAS21ptp, in the host galaxy NGC~7828 \citep[$z = 0.01911$, Tully-Fisher distance modulus $\mu=34.71$ mag; ][through NED\footnote{The NASA/IPAC Extragalactic Database (NED), \href{http://ned.ipac.caltech.edu/}{http://ned.ipac.caltech.edu/}}]{theureau07}. The host galaxy is interacting with a smaller galaxy NGC~7829, forming the ring galaxy system Arp 144.
No meaningful pre-discovery upper limit exists as the object was emerging from solar conjunction at the time of discovery.
The transient was subsequently reported to the Transient Name Server (TNS\footnote{\href{https://www.wis-tns.org/object/2021ocs}{https://www.wis-tns.org/object/2021ocs}}) by the Zwicky Transient Facility \citep[ZTF,][]{bellm19} and Pan-STARRS1 \citep{chambers16} surveys as ZTF21abhrpia and PS21hlk, respectively. 
Spectral classification reported to the TNS suggests that SN 2021ocs is a type-Ic SN around one to two weeks after maximum light \citep{huber21}. 
Here we report additional spectral observations of SN 2021ocs obtained at late time (148 d after light curve peak)\footnote{{The spectrum and light curves are available as the Data behind the Figure (DbF).}}. 

\section{Observations and Data Reduction}

The classification spectrum of SN~2021ocs was obtained on 2021-06-13 using the SNIFS spectrograph \citep{lantz04} at the 2.2 m University of Hawaii telescope on Mauna Kea, as part of the Spectroscopic Classification of Astronomical Transients (SCAT) survey \citep{tucker22}.
Late-time spectroscopy of SN~2021ocs was conducted on the night of 2021-10-26, as part of the FORS+ Survey of Supernovae in Late Times program (\textit{FOSSIL}, Kuncarayakti et al. in prep.), using the FORS2 instrument \citep{appenzeller98} attached to the ESO Very Large Telescope (VLT) at Cerro Paranal Observatory, Chile. 
{FOSSIL targets all observable CCSNe brighter than $\sim18.5$ mag in the photospheric phase, which are expected to be $\sim$21-22 mag or brighter when observed in the nebular phase, with the goal of obtaining nebular spectra for a large sample of objects in a magnitude-limited, unbiased way.}

We used FORS2 with grism 300V and the 1.6" slit, achieving a wide wavelength coverage of 3500--9500~\AA~and a spectral resolution of $R\sim400$ measured from the narrow sky emission lines. The sky conditions were photometric with seeing around 0.5" during the length of the integration. 
The spectroscopic observations were obtained with $2 \times 1300$~s exposures, accompanied with 20~s \textit{g}-band imaging under 0.75" seeing conditions. Spectrophotometric standard stars were observed using the same grism setting. The data were reduced using the ESOReflex \citep{freudling13} pipeline following standard procedures. The excellent seeing conditions allowed reasonable background subtraction during the spectrum extraction, evidenced by the absence of narrow host galaxy emission lines.

The \textit{g}-band photometry of FORS2 was measured with our own PSF photometry code. Synthetic photometry was performed on the nebular spectrum to derive the \textit{g} and \textit{r} synthetic magnitudes; the obtained $(g-r)$ color was then applied to the \textit{g}-band photometry from imaging to produce an \textit{r}-band magnitude to be used in the light curve.
In addition to the FORS2 \textit{gr} imaging, photometry was obtained from a number of sources. Public ATLAS forced photometry in the \textit{o} and \textit{c} bands was obtained from the ATLAS Forced Photometry server\footnote{\href{https://fallingstar-data.com/forcedphot/}{https://fallingstar-data.com/forcedphot/}}, and ZTF \textit{g} and \textit{r}-band photometry through the 
{ZTF forced photometry service\footnote{\url{https://ztfweb.ipac.caltech.edu/cgi-bin/requestForcedPhotometry.cgi}} \citep{masci19}}.
Forced photometry data from the PanSTARRS1 survey are used, yielding \textit{i} and \textit{w}-band photometry.


\section{Results and Discussion}

\subsection{Early spectrum and light curve}

\begin{figure}
\centering
\includegraphics[width=\linewidth]{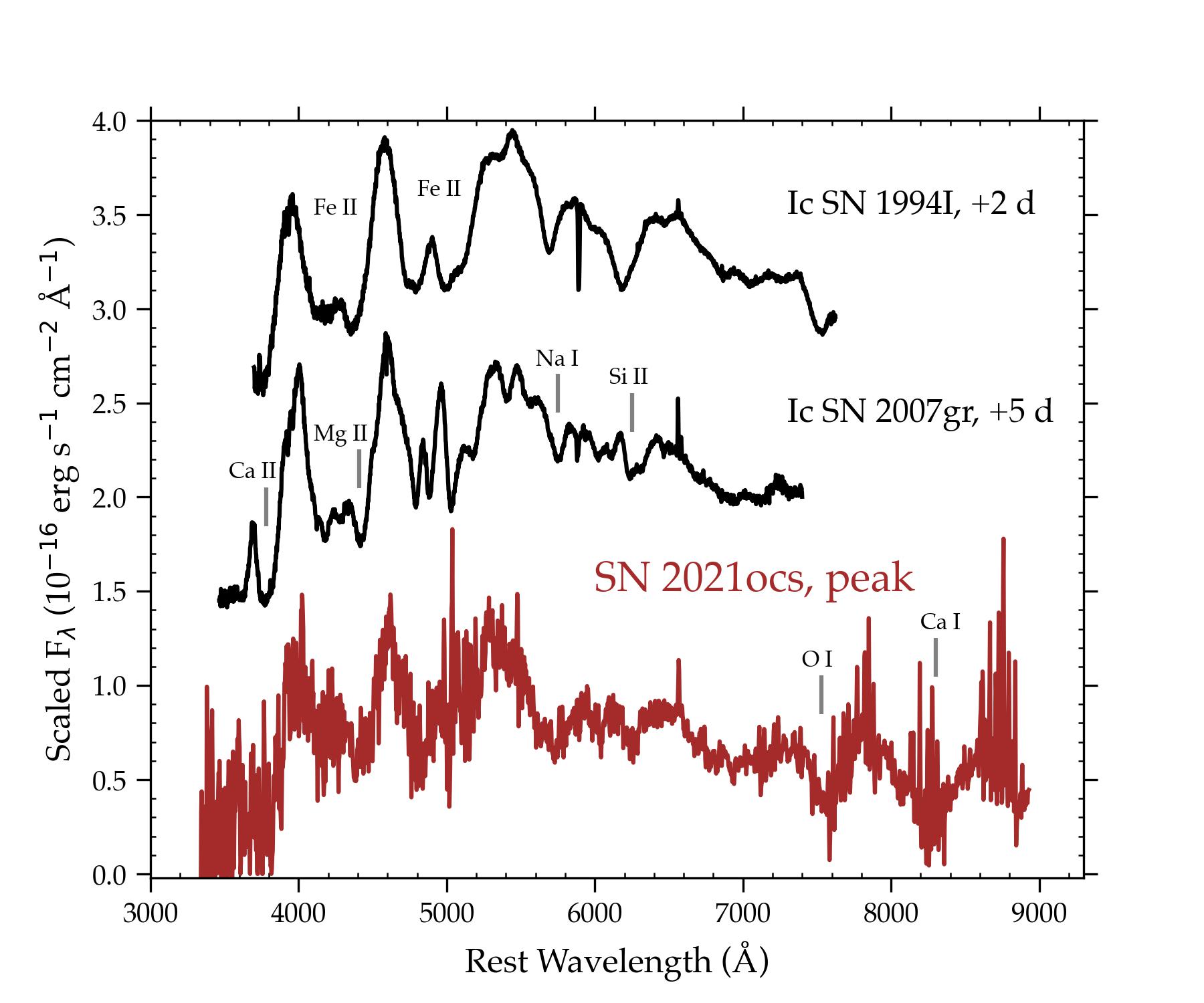} 
\caption{Classification spectrum of SN 2021ocs obtained shortly after the light curve peak, compared to those of well-observed SNe Ic at similar epochs. The spectra and phases ({defined as time relative to the light curve peak throughout the paper}) of the comparison objects were taken from the WISeREP repository \citep{yaron12}, originally from \citet{modjaz14}. Prominent absorption features typical of SN type Ic are indicated.}
\label{specearly}
\end{figure}

The classification of SN 2021ocs as a type-Ic SN around the light curve peak \citep{huber21} was obtained using the SNID tool \citep{blondin07}. Figure~\ref{specearly} shows a comparison between SN 2021ocs and the well-studied type-Ic SNe, SN~1994I and SN~2007gr, as two of the best matches obtained by SNID. SN 2021ocs appears similar to normal type-Ic SNe around the light curve peak. The lack of narrow Na I D $\lambda\lambda5889,5896$ absorption line in the spectrum (cf. SN 1994I where the narrow absorption line is strong, and weaker in SN 2007gr) suggests that the amount of intervening extinction is minimal. Henceforth, we assume no host galaxy extinction for SN 2021ocs. The reported foreground extinction for NGC~7828 is similarly negligible as it is at the level of the photometric uncertainty, $A_V = 0.086$~mag \citep[][via NED]{schlafly11}.

\begin{figure*}
\gridline{\fig{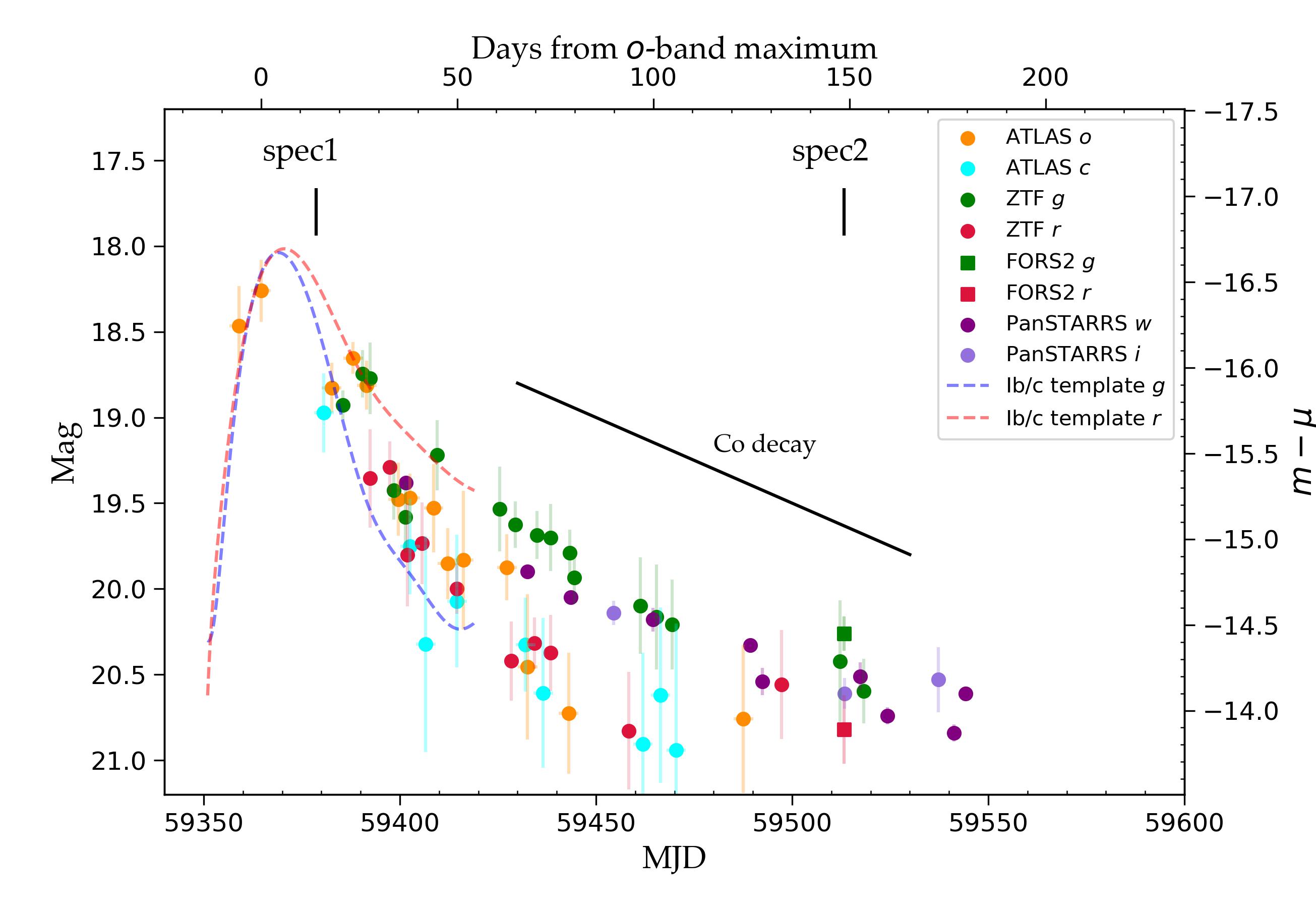}{0.62\textwidth}{(a)}
          \fig{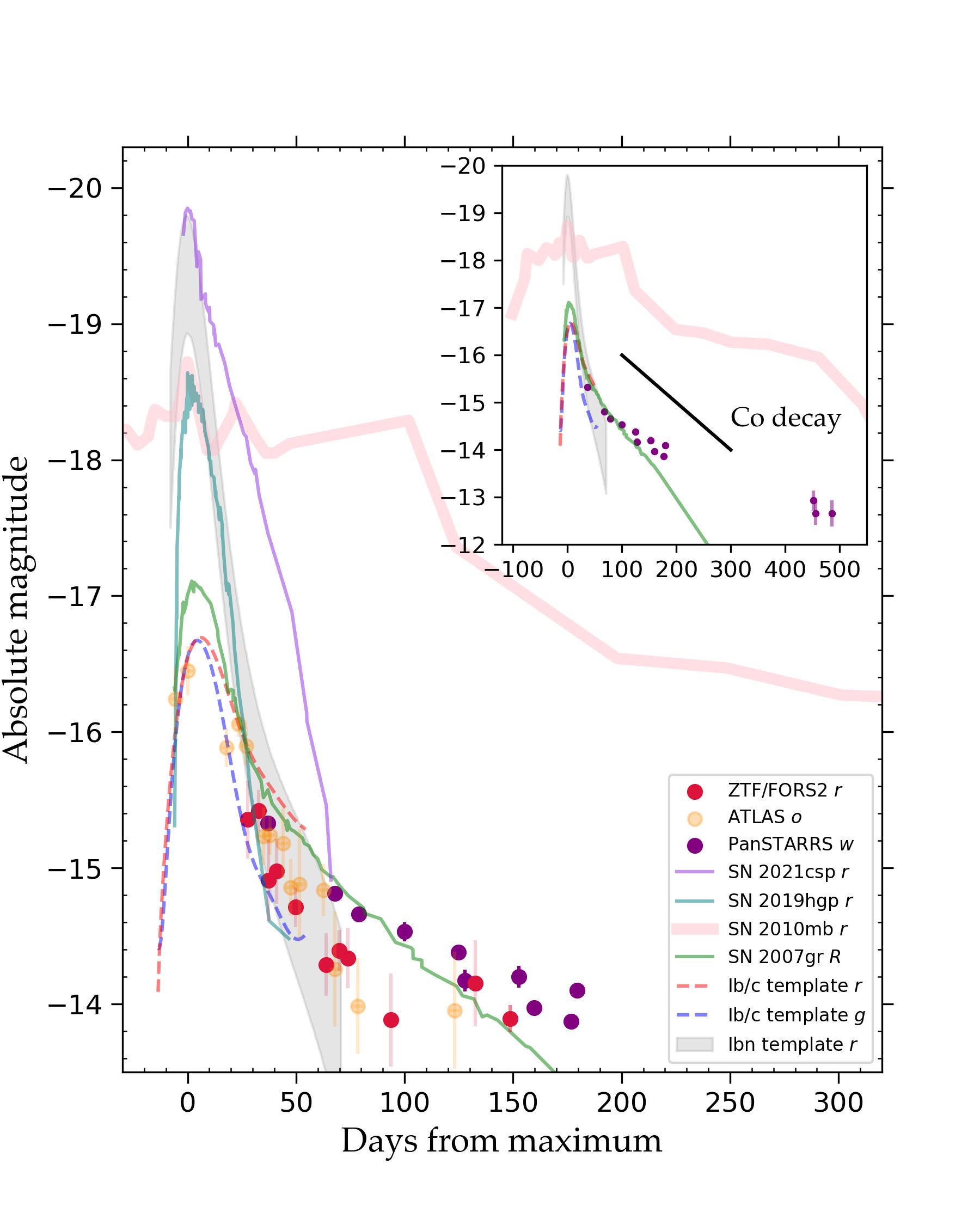}{0.35\textwidth}{(b)}
          }
\caption{(a) Light curves of SN 2021ocs in ATLAS \textit{o} (orange symbols) and \textit{c} (cyan) with 5-{day} binning, {ZTF 3-day binned 5$\sigma$ detections and FORS2 in \textit{g} (green symbols) and \textit{r} (red)}, and PanSTARRS \textit{w} (purple) and \textit{i} (light purple) bands. 
The template SN Ib/c light curves from \citet{taddia15} are plotted in dashed lines ({blue} for \textit{g} and red for \textit{r}). The epochs of the first (classification) and second (nebular) spectra are indicated, along with the radioactive $^{56}$Co decay rate assuming complete $\gamma$-ray trapping (0.01 mag d$^{-1}$). The ATLAS \textit{c} bandpass corresponds roughly to \textit{g}+\textit{r} bands, while \textit{o} covers \textit{r}+\textit{i}. The PanSTARRS \textit{w} band is a white-light bandpass covering \textit{gri} bands.
(b) Absolute-magnitude light curve comparison against well-observed type-Icn SNe 2010mb \citep{benami14}, 2019hgp \citep{galyam22}, and 2021csp \citep{perley22}. SN~2010mb is a peculiar case of Icn-related objects with its broad light curve. A light curve template of SNe Ibn \citep{hosseinzadeh17} is also plotted, {alongside the Ib/c templates and the $R$-band light curve of SN~2007gr \citep[$\mu=29.84$ mag,][]{hunter09}}. Inset shows a wider phase range covering PanSTARRS late detections.
}
\label{lcurve}
\end{figure*}

While the light curves of SN 2021ocs are not very well sampled, the earliest data points from ATLAS suggest that it was rising in brightness (Figure~\ref{lcurve}a), and the subsequent epochs indicate that it {declined steadily, as is typical for most SESN light curves, though with slight fluctuations.}
Comparing to the templates of early-time light curves of SNe Ib/c \citep{taddia15}, it appears that the peak {of the} light curve occurred around a week after the brightest point in the photometry, which implies that the classification spectrum was taken within 1-2 weeks from the peak brightness, consistent with the phase determined from spectral matching. 
The comparison between the light curves and the templates suggests that SN 2021ocs peaked around $M \sim -16.7$~mag, which is fainter than most SESNe, but still within the range of the previously observed objects  {(\citealt{taddia15,zheng22}, also see e.g. \citealt{perley20}, Figure~7).}
{Alternatively, the peak could have occurred during the solar conjunction before the first detections, which would imply that the light curve could be broader and more luminous. However, in this case the classification spectrum would suggest a considerably older phase. This alternative scenario cannot be ruled out, but requires strong additional assumptions on the early spectral evolution in order to be consistent with the classification spectrum.}

In Figure~\ref{lcurve}b, the light curves of SN~2021ocs are compared with 
{those of SN~2007gr\footnote{Obtained from the Open Astronomy Catalog API, \url{https://github.com/astrocatalogs/OACAPI.}} and H-poor interacting SNe} (see section \ref{interpretation}). Assuming typical early SN Ic evolution, SN~2021ocs shows a gradual decline after the light curve peak. In general, SN~2021ocs shows slower decline compared to the Ibn/Icn population, although there exist individual objects that exhibit significantly broader light curves \citep[e.g.][]{benami14,kool21}.

{While the light curve peak brightness and width are unlikely to diverge significantly compared to the bulk of SESNe, SN~2021ocs appears to be peculiarly blue from the first detections in $g$ and $r$ bands up until late phases\footnote{Note that there is no $r$-band detection in the ZTF public data stream, see e.g. \url{https://alerce.online/object/ZTF21abhrpia}.}.}
The host-subtracted photometry from ZTF, and the FORS2 photometry, indicate $(g-r) \lesssim -0.5$~mag {throughout the evolution} following the light curve peak. Such a blue color is not normally seen in regular SESNe at epochs post peak. Indeed, even {at} 2-3 weeks before maximum light, such a blue color is rare, and the color index usually stays $>0$~mag throughout the SN evolution \citep[e.g.][]{taddia15,zheng22}. 

During the decay phase, the light curves show possible {flattening leading to a tail phase slope} shallower than the $^{56}$Co decay rate {assuming complete $\gamma$-ray trapping} (Figure~\ref{lcurve}b, inset). 
A decay rate slower than $^{56}$Co decay indicates that the light curve is not powered solely by radioactive decay, and thus may be powered also by other processes e.g. late-time ejecta-CSM interaction \citep[e.g.][]{maeda15,dessart22} or magnetar spin-down \citep[e.g.][]{afsariardchi21}.
{The blue color is reminiscent of the interaction-powered type-Ibn SNe \citep[see e.g.][Figure 12]{ho21}, suggesting that a similar powering mechanism is likely to be at play.}


\subsection{Nebular spectrum}

\begin{figure}
\centering
\includegraphics[width=\linewidth]{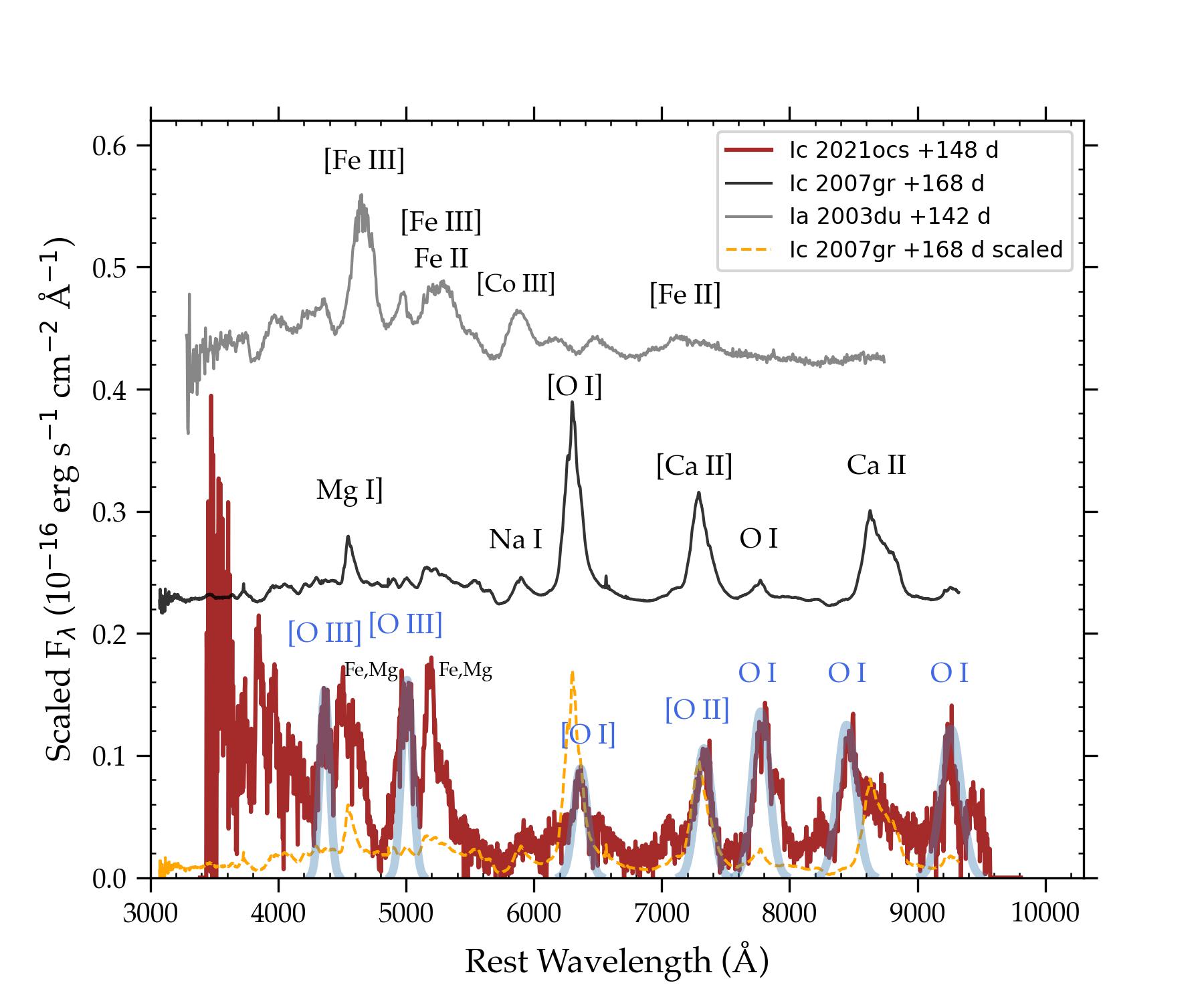} 
\caption{Nebular spectrum of SN~2021ocs (red), compared to typical nebular spectra of SNe Ic and Ia at similar epochs. The comparison spectra are of SNe 2007gr \citep[black, ][]{shivvers19} and 2003du \citep[grey, ][]{stanishev07}, taken from WISeREP. The spectrum of SN~2021ocs shows a number of O lines with superposed model lines, assuming a Gaussian profile with ${\rm FWHM} = 6000 {\rm ~km~s}^{-1}$ (light blue). The flux of the spectrum of SN~2007gr is scaled to match the $r$-band magnitude of SN~2021ocs at +148 d; the spectrum is plotted twice for clarity of the comparison.}
\label{nebspec1}
\end{figure}

The nebular spectrum of SN 2021ocs, obtained +148~d after the \textit{o}-band peak, shows characteristics not regularly seen in an SESN nebular spectrum. While the latter is typically dominated by strong emission lines of [O~I] $\lambda\lambda6300,6364$ doublet, [Ca~II] $\lambda\lambda7292,7324$ doublet, and Ca~II $\lambda\lambda 8498,8542,8662$ triplet, SN~2021ocs shows a spectrum with more than five emission lines of similar strength across the spectrum (Figure~\ref{nebspec1}). 
{Such a spectrum has never been seen among $\sim200$ SESN nebular spectra in the literature \citep{fang22,prentice22,tauben09}, suggesting a very rare occurrence rate of under $1\%$.}
It is also markedly different compared to the nebular spectra of type-Ia SNe, which are dominated by Fe-peak elements \citep[e.g.][]{tauben13}.
Strong H emission lines are absent in SN~2021ocs, which supports the initial SESN classification. The emission lines in the spectrum may be attributed to different transitions and ionization states of O, i.e. O~I, [O~I], [O~II], and [O~III], with similar line widths of $\sim6000$~km~s$^{-1}$ (FWHM, full-width at half-maximum). The commonly seen [O~I] $\lambda\lambda6300,6364$ doublet is present, possibly superposed on a broader base extending to $\pm12~000$~km~s$^{-1}$ {which may be attributed to Fe~II \citep{dessart21}}. [O~II] $\lambda\lambda7320,7330$ is present and likely to be blended with the commonly seen [Ca~II] $\lambda\lambda7292,7324$. It is clear that the [O~I] $\lambda\lambda6300,6364$ and [Ca~II] $\lambda\lambda7292,7324$ lines are not the strongest lines in the nebular spectrum, which is atypical for SESNe. Furthermore, the flux ratio of [O~I] $\lambda\lambda6300,6364$ to [Ca~II] $\lambda\lambda7292,7324$ is typically $>1$ for type-Ic SNe \citep[e.g.][]{fang22}, which is not the case in SN~2021ocs.


In the red part of the spectrum, three permitted O~I emission lines are seen at $\lambda\lambda$7774, 8446, and 9263\footnote{These lines are themselves multiplets of several oxygen transitions.}. While the O~I $\lambda$7774 is weak in SESN nebular spectra, it is exceptionally strong in SN~2021ocs, {suggesting high density conditions}. This line is possibly contaminated by Mg~I $\lambda\lambda$7877, 7896 in the red shoulder (Figure~\ref{nebspec2}).
The $\lambda$8446 line is similarly contaminated by the broad Ca~II triplet on the red side (see comparison with scaled SN~2007gr in Figure~\ref{nebspec1}), although the peak is still clearly prominent. The [O~I] $\lambda$9263 line is blended with Mg~II $\lambda9224$.
The peak intensities of the [O~I] $\lambda$8446 and $\lambda$9263 lines are similar to {the O~I $\lambda$7774 line}. In typical SESNe, these two redder lines are either very weak or missing. 
These three O~I lines are considered as the most persistent lines of oxygen in the optical and near-infrared regimes, as they appear over a broad range of conditions in spectroscopic experiments \citep{sansonetti05}.

In the blue, two {broad} [O~III] emission lines are seen, the $\lambda\lambda$4959, 5007 doublet and $\lambda$4363. They are accompanied with broader emission lines on the red side ($\sim$ 4600~\AA, 5200~\AA), possibly arising from the Fe~I/Fe~II complexes \citep[see e.g. Fig. 7 of][]{dessart21}, with prominent peaks of Mg~I and Mg~II at $\lambda\lambda$ 4481, 4571, and 5170. 
Broad [O~III] at late phase is also seen in a number of interacting SESNe, e.g. SN~1993J \citep{matheson00} and SN~2014C \citep{milisav15}.
Mg appears conspicuously in the spectrum of SN~2021ocs, with other lines at $\lambda\lambda$ 3832 (triplet), 8224 (possible blend with [O~I] $\lambda$8221), and 9436 (doublet). The Mg emission lines are seen weakly also in some SNe Ibn such as SN~2006jc \citep{pastorello07}.

\subsection{Interpretation: CSM interaction or a pulsar wind nebula?}
\label{interpretation}

\begin{figure*}
\centering
\includegraphics[width=\linewidth]{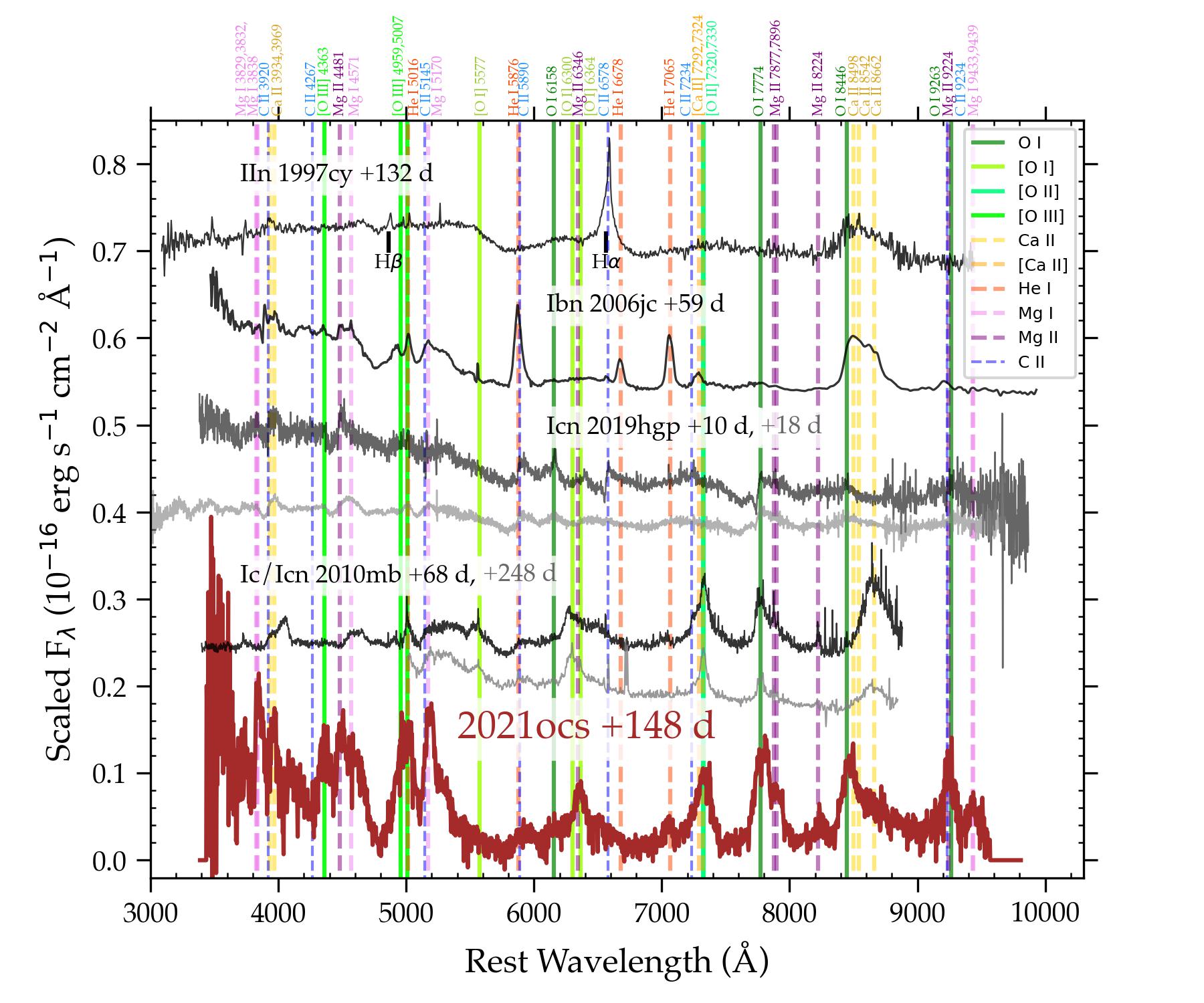} 
\caption{Comparison of the SN~2021ocs spectrum with interacting SNe of types IIn (SN~1997cy, \citealt{turatto00}), Ibn (SN~2006jc, \citealt{pastorello07}), and Icn (SN~2019hgp, \citealt{galyam22}, and SN~2010mb, \citealt{benami14}). The epochs are relative to maximum light.
Spectra were taken from WISeREP. Significant emission lines are identified.
}
\label{nebspec2}
\end{figure*}

Comparing the nebular spectrum of SN~2021ocs to a number of interacting SNe yields few similarities while the spectrum remains unique and unparalleled in the literature. 
{The blue color and light curve flattening in SN~2021ocs may be explained by CSM interaction, which warrants a spectral comparison with interacting SNe.}
Figure~\ref{nebspec2} shows SN~2021ocs spectrum compared to interacting SNe of types IIn, Ibn, and Icn which have relatively good coverage in late phases\footnote{Some objects (i.e. SNe 2006jc and 2019hgp in this case) evolve more rapidly and thus the phases in days are shorter compared to the slower-evolving objects.}. 

The strong narrow H emission lines defining the type-IIn SNe \citep{schlegel90} are clearly absent in SN~2021ocs. SNe Ibn show prominent narrow He emission lines and a rising blue part of the spectrum \citep{pastorello07}, which originates from Fe emission \citep{dessart21,dessart22}. Similarly this blue rise is also seen in SN~2021ocs, while strong He lines are absent except for the weak He~I $\lambda7065$, and $\lambda$5876 which is blended with the Na~I doublet $\lambda\lambda$5890,5896. Also, He~I $\lambda$5016 may be contaminating the [O~III] doublet $\lambda\lambda$4959,5007. The weak He lines in the spectrum of SN~2021ocs suggest that little He is present.

Similar to SNe IIn and Ibn, SNe Icn \citep{galyam22} also show the Fe bumps at late time. In SNe Icn, the initially strong lines of ionized C disappear at later phases to give way to nebular O~I lines in the red part of the spectrum. These O~I lines are present in SN~2021ocs, which may suggest a connection to SNe Icn which are interpreted as explosions of Wolf-Rayet (WR) stars within a H/He-deficient CSM.
Considering the spectra and light curves, it is possible that SN~2021ocs fits this scenario as well, although the CSM interaction did not occur immediately after the explosion. The expanding ejecta interacted with H/He-poor CSM and the interaction drove a reverse shock that ionized the O/Mg-rich outer part of the ejecta. Clumping and inhomogeneity in the ejecta, CSM, and also in the $^{56}$Ni distribution, would cause different levels of compression and ionization stages in the gas, which could give rise to the various ionization states seen in the O and Mg emission lines.
The absence/weakness of C and He lines sets apart SN~2021ocs from the general population of SN Ibn and Icn.

\begin{figure*}
\centering
\includegraphics[width=\linewidth]{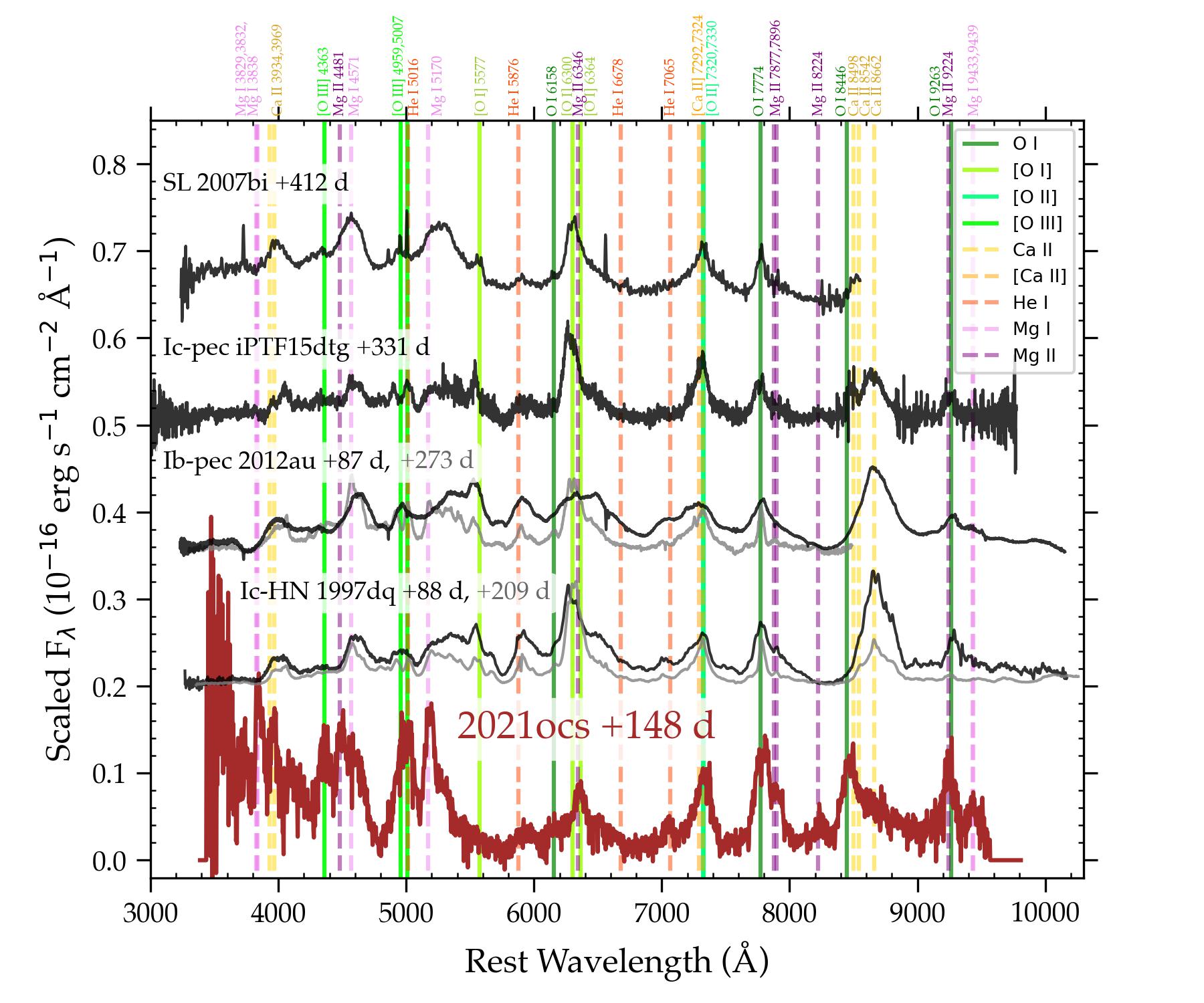} 
\caption{Comparison of the SN~2021ocs spectrum with SNe harboring a pulsar/magnetar wind nebula in the sample of \citet{milisav18}: SN~2007bi (SLSN-I, \citealt{galyam09}), iPTF15dtg (Ic with broad light curves, \citealt{taddia19}), SN~2012au (Ib-peculiar, \citealt{milisav18}), and SN~1997dq (Ic-hypernova, \citealt{matheson01,tauben09}). The epochs are relative to maximum light.
Spectra were taken from WISeREP. Significant emission lines are identified.
}
\label{nebspec3}
\end{figure*}

Broad [O~III] $\lambda\lambda4959,5007$ emission is usually only seen in very late times in CCSNe, a few years after the explosion, and has been interpreted as an evidence of a pulsar/magnetar wind nebula \citep{chevalier92,milisav18} {alternatively to CSM interaction}. In this case, the line is also accompanied by [O~I] and [O~II], which is interpreted as different ionization layers by photoionization.
The [O~III] line is seen in a variety of SESNe including the normal and superluminous ones \citep{milisav18}. This subset of objects which show broad late-time [O~III] curiously display a narrow ($\sim2000$~km~s$^{-1}$) and relatively strong O~I $\lambda$7774 line in the nebular phase around a half to one year post explosion, although in these cases it is neither strong nor accompanied by the same set of lines seen in SN~2021ocs.
Figure~\ref{nebspec3} shows the nebular spectrum of SN~2021ocs compared to such objects. Generally the agreement is poor: while they show similar line profiles in [O~II] $\lambda\lambda7320,7330$ (sloping blue shoulder, contamination by [Ca~II]) and O~I $\lambda\lambda$7774 (sloping red shoulder, contamination by Mg II), striking differences are seen in the O~I $\lambda$8446 line, Mg lines redward of 8000 \AA, and in the various strong lines in the blue, which all are absent or very weak in the comparison objects.
The association of SN~2021ocs with a wind nebula caused by a magnetized central object is therefore weak, and CSM interaction remains as our preferred interpretation of the observed properties in the spectra and light curves. Drawing analogy with type-Ibn/Icn SNe \citep[e.g.][]{pastorello07,benami14,ho21,galyam22}, SN~2021ocs shows a {similarly blue $(g-r)$ color and a rising blue continuum possibly extending to the ultraviolet \citep[cf. interaction models of][]{dessart22uv}}, and prominent emission lines of O and Mg, suggesting similar mechanism of CSM interaction.
Even with CSM interaction, SN~2021ocs appears to be underluminous ($M_o = -16.7$ mag, compared to $M_R = -17.9 \pm 0.7$ mag for SNe Ic in the sample of \citealt{zheng22}). This suggests a small amount of $^{56}$Ni, which is another similarity to SNe Ibn and Icn.

In SN~2021ocs, the CSM interaction likely occurred past the light curve peak.
Clearly, if interaction is present, it cannot be strong early on.
A delayed interaction may be interpreted as a detached or low-density nearby CSM, which was surrounding the SN progenitor star at the time of explosion. 
Furthermore, for the interaction to be dominating at late times, the amount of CSM must be {small} \citep[see e.g.][]{dessart22}.
Assuming an ejecta velocity of 10~000 km~s$^{-1}$, if CSM interaction {started $\sim$50~d after the explosion as inferred from the blue $(g-r)$ color,} then the distance traversed by the unimpeded SN ejecta {would be $\sim4 \times 10^{15}$ cm from the progenitor star}. {This inferred CSM distance is similar to those in other interacting SESNe with detached CSM (e.g. SN~2017dio, \citealt{hk18})}.
In the case of SN~2021ocs, as a type-Ic SN its progenitor could have been a carbon-oxygen star with a wind velocity of $\sim$1000 km~s$^{-1}$. If the CSM was formed through such a wind, the mass loss probed by the CSM interaction would have occurred $\sim$500 d before the SN. At this time, the progenitor star {would have just finished} the C-shell burning stage and {starting} Ne burning \citep{fuller17}.
{The short CSM distance and time before the SN explosion indicate} that the SN progenitor star could have been surrounded by CSM at the time of explosion although not as embedded as in the case of SNe Icn/Ibn, which show a rapidly increasing CSM density toward the SN progenitor \citep[steeper than $r^{-2}$,][]{maeda22}.
The immediate circumstellar environment of SN~2021ocs, therefore, was relatively clean compared to those of SNe Icn/Ibn, as the explosion was first seen as a normal type-Ic SN without CSM interaction. This suggests that the CSM density was low closer to the progenitor star, {implying a possible CSM distribution in a detached torus or disk, clumps, or a shell, which in any case contains a central cavity}. 
Two possible interpretations arise regarding the progenitor mass loss as it approaches the terminal explosion: (1) a low progenitor mass-loss rate, suggesting that mass loss may have become weaker once the stripping reached the inner O-Mg core, or (2) the progenitor ejected some material which  pushed away slower CSM in the vicinity, creating a cavity. In either case, SN~2021ocs represents the most extreme case of envelope stripping where the O-Mg layer of the progenitor is exposed. Within the {massive star} SN ejecta-CSM interaction case, it is positioned at the end of the sequence of CSM build-up resulting from the progenitor stripping: IIn $\rightarrow$ Ibn $\rightarrow$ Icn $\rightarrow$ SN~2021ocs.   



%
%

\section{Summary}

This Letter presents photometry and spectroscopy of SN~2021ocs, which yield a peculiar nebular spectrum dominated by O and Mg emission lines. The unique set of emission lines, blue continuum and color, and slowly-declining light curve tail suggest that interaction with a H/He-poor CSM took place in SN~2021ocs. The absence of signs of interaction in the early spectrum suggests that the CSM density close to the progenitor star was low. Comparing with interacting SESNe of types Ibn and Icn, SN~2021ocs appears to be more stripped as the deep O-Mg layer in the progenitor is exposed, and the outer C-O and He-rich layers are stripped away. SN~2021ocs poses yet another challenge to stellar evolution theory, regarding the final phases of evolution of massive stars.
{With its unique spectral and photometric behavior, it represents a rare case not previously considered. Modeling the evolutionary pathways, mass loss, and explosion of a highly stripped star with an exposed O-Mg layer will provide insights in comparison with the observations. Future observations of transients should consider targets with unusual colors, albeit a spectroscopically normal appearance, in order to uncover similar peculiar objects.}

\begin{acknowledgments}
The anonymous referee is thanked for the useful feedback which substantially improved the paper.
Stephen Smartt, Masaomi Tanaka, and Lucy McNeill are thanked for the discussions and suggestions on the paper.
H.K. and T.N. were funded by the Academy of Finland projects 324504 and 328898.  
The work is supported by the JSPS Open Partnership Bilateral Joint Research Project between Japan and Finland (JPJSBP120229923). K.M. acknowledges support from the Japan Society for the Promotion of Science (JSPS) KAKENHI grant JP18H05223 and JP20H00174.
SM acknowledges support from the Academy of Finland project 350458.
This work has made use of data from the Asteroid Terrestrial-impact Last Alert System (ATLAS) project. The Asteroid Terrestrial-impact Last Alert System (ATLAS) project is primarily funded to search for near earth asteroids through NASA grants NN12AR55G, 80NSSC18K0284, and 80NSSC18K1575; byproducts of the NEO search include images and catalogs from the survey area. This work was partially funded by Kepler/K2 grant J1944/80NSSC19K0112 and HST GO-15889, and STFC grants ST/T000198/1 and ST/S006109/1. The ATLAS science products have been made possible through the contributions of the University of Hawaii Institute for Astronomy, the Queen’s University Belfast, the Space Telescope Science Institute, the South African Astronomical Observatory, and The Millennium Institute of Astrophysics (MAS), Chile.
The Pan-STARRS1 Surveys (PS1) and the PS1 public science archive have been made possible through contributions by the Institute for Astronomy, the University of Hawaii, the Pan-STARRS Project Office, the Max-Planck Society and its participating institutes, the Max Planck Institute for Astronomy, Heidelberg and the Max Planck Institute for Extraterrestrial Physics, Garching, The Johns Hopkins University, Durham University, the University of Edinburgh, the Queen's University Belfast, the Harvard-Smithsonian Center for Astrophysics, the Las Cumbres Observatory Global Telescope Network Incorporated, the National Central University of Taiwan, the Space Telescope Science Institute, the National Aeronautics and Space Administration under Grant No. NNX08AR22G issued through the Planetary Science Division of the NASA Science Mission Directorate, the National Science Foundation Grant No. AST-1238877, the University of Maryland, Eotvos Lorand University (ELTE), the Los Alamos National Laboratory, and the Gordon and Betty Moore Foundation.
The ZTF forced-photometry service was funded under the Heising-Simons Foundation grant $\#$12540303 (PI: Graham).
This work was funded by ANID, Millennium Science Initiative, ICN12\_009.
\end{acknowledgments}

%

\vspace{5mm}
\facilities{VLT:Antu(FORS2), UH:2.2m(SNIFS), PS1}


\software{IRAF \citep{tody86,tody93}, ESOReflex \citep{freudling13}}

\bibliography{bib21ocs}{}
\bibliographystyle{aasjournal}

%



\end{document}